\documentclass[smallabstract,smallcaptions]{dccpaper}
\usepackage{epsfig}
\usepackage{citesort}
\usepackage{amsmath}
\usepackage{amssymb}
\usepackage{color}
\usepackage{url}
\usepackage{booktabs}
\usepackage{enumerate}
\usepackage{float}
\usepackage{titlesec}

\titlespacing\section{0pt}{12pt plus 4pt minus 2pt}{0pt plus 2pt minus 2pt}
\newlength{\figurewidth}
\newlength{\smallfigurewidth}
\usepackage{setspace}

\begin{document}

\setcounter{page}{1}
\pagenumbering{arabic}

\title{\large \textbf{Semantic Perceptual Image Compression using Deep Convolution Networks}}
 \author{%
     Aaditya Prakash, Nick Moran, Solomon Garber, Antonella DiLillo and James Storer\\[0.5em]
     {\small\begin{minipage}{\linewidth}\begin{center}
     \begin{tabular}{ccc}
     Brandeis University \\
     \url{{aprakash,nemtiax,solomongarber,dilant,storer}@brandeis.edu}  \\
     \\
     \end{tabular}
     \end{center}\end{minipage}}
}
\maketitle
\longpage

\begin{abstract}
It has long been considered a significant problem to improve the visual quality of lossy image and video compression.
Recent advances in computing power together with the availability of large
training data sets has increased interest in the application of deep learning \textsc{cnn}s to address image recognition and image processing tasks.
Here, we present a powerful \textsc{cnn} tailored to the specific task of semantic image understanding to achieve higher visual quality in lossy compression.
A modest increase in complexity is incorporated to the encoder which allows a standard, off-the-shelf \textsc{jpeg} decoder to be used.
While \textsc{jpeg} encoding may be optimized for generic images, the process is ultimately unaware of the specific content of the image to be compressed.
Our technique makes \textsc{jpeg} content-aware by designing and training a model to identify multiple semantic regions in a given image.
Unlike object detection techniques, our model does not require labeling of object positions and is able to identify objects in a single pass.
We present a new \textsc{cnn} architecture directed specifically to image compression, which generates a map that highlights semantically-salient regions so that they can be encoded at higher quality as compared to background regions.
By adding a complete set of features for every class, and then taking a threshold over the sum of all feature activations, we generate a map that highlights semantically-salient regions so that they can be encoded at a better quality compared to background regions.
Experiments are presented on the Kodak PhotoCD dataset and the MIT Saliency Benchmark dataset, in which our algorithm achieves higher visual quality for the same compressed size.\footnote{Code for the model is available at https://github.com/iamaaditya/image-compression-cnn}

\end{abstract}

\section{Introduction and Related Work}

We employ a Convolutional Neural Network (\textsc{cnn}) tailored to the specific task of semantic image understanding to achieve higher visual quality in lossy image compression.
We focus on the \textsc{jpeg} standard, which remains the dominant image representation on the internet and in consumer electronics. Several attempts have been made to improve upon its lossy image compression, for example WebP \cite{ginesu2012objective} and Residual-GRU \cite{toderici2016full}, but many of these require custom decoders and are not sufficiently content-aware.

\begin{figure}
    \centering
    \includegraphics[scale=0.34]{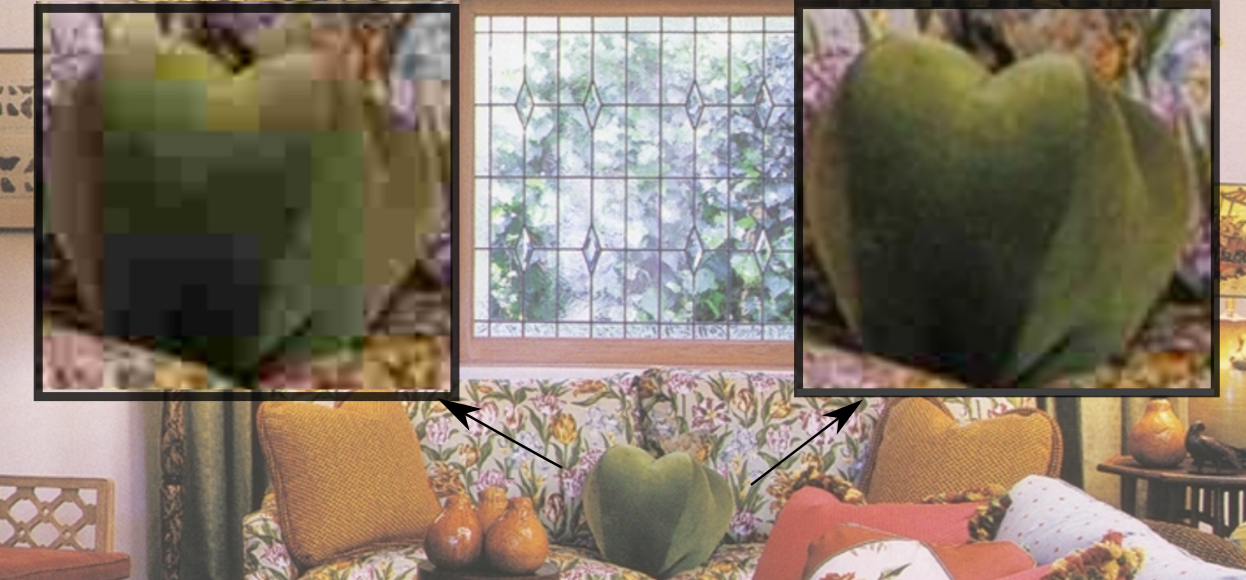}
    \caption{Comparison of compression of semantic objects in standard \textsc{jpeg}[left] and our model [right]\label{fig_compressing_demo}}
\end{figure}

We improve the visual quality of standard \textsc{jpeg} by using a higher bit rate to encode image regions flagged by our model as containing content of interest and lowering the bit rate elsewhere in the image. 
With our enhanced \textsc{jpeg} encoder, the quantization of each region is informed by knowledge of the image content.
Human vision naturally focuses on familiar objects, and is particularly sensitive to distortions of these objects as compared to distortions of background details \cite{jiang2015salicon}.
By improving the signal-to-noise ratio within multiple regions of interest, we improve visual quality of those regions, while preserving overall \texttt{PSNR} and compression ratio.
A second encoding pass produces a final \textsc{jpeg} encoding that may be decoded with any standard off-the-shelf \textsc{jpeg} decoder.
Measuring visual quality is an ongoing area of research and there is no consensus among researchers on the proper metric.
Yuri et al \cite{kerofsky2015perceptual} showed that \texttt{PSNR} has severe limitations as an image comparison metric.
Richter et al \cite{richter2009ms}\cite{richter2011ssim} addressed structural similarity (\texttt{SSIM}\cite{ssim} and \texttt{MS-SSIM}\cite{msssim}) for \textsc{jpeg} and \textsc{jpeg} 2000.
We evaluate on these metrics, as well as \texttt{VIFP}\cite{vifp},  \texttt{PSNR-HVS}\cite{psnrhvs} and \texttt{PSNR-HVSM}\cite{psnrhvsm}, which have been shown to correlate with subjective visual quality.  Figure~\ref{fig_compressing_demo} compares close-up views of a salient object in a standard \textsc{jpeg} and our new content-aware method.

\textsc{cnn}s have been successfully applied to a variety of computer vision tasks \cite{krizhevsky2012imagenet}.
The feature extraction and transfer learning capabilities of \textsc{cnn}s are well known \cite{zeiler2014visualizing}, as are their ability to classify images by their most prominent object \cite{he2015deep}, and compute a bounding box \cite{girshick2014rich}.
Some success has been obtained in predicting the visual saliency map of a given image \cite{jiang2015salicon},\cite{kummerer2014deep}.
Previous work has shown that semantic object detection has a variety of advantages over saliency maps \cite{mnih2014recurrent}, \cite{zund2013content}.
Semantic detection recognizes discrete objects and is thus able to generate maps that are more coherent for human perception.
Visual saliency models are based on human eye fixations, and thus produce results which do not capture object boundaries \cite{kummerer2014deep}.
This is evident in the results obtained by Stella et al \cite{stella2009image}, in which image compression is guided by a multi-scale saliency map, and the obtained images show blurred edges and soft focus.

We present a \textsc{cnn} designed to locate multiple regions of interest (\textsc{roi}) within a single image.
Our model differs from traditional object detection models like \cite{dai2016r}, \cite{girshick2014rich} as these models are restricted to detecting a single salient object in an image.
It captures the structure of the depicted scene and thus maintains the integrity of semantic objects, unlike results produced using human eye fixations \cite{liu2015predicting}.
We produce a single class-invariant feature map by learning separate feature maps for each of a set of object classes and then summing over the top features.
Because this task does not require precise identification of object boundaries, our system is able to capture multiple salient regions of an image in a single pass, as opposed to standard object detection \textsc{cnn}s, which require multiple passes over the image to identify and locate multiple objects.
Model training need only be done offline, and encoding with our model employs a standard \textsc{jpeg} encoder combined with efficient computation of saliency maps (over 90 images per second using a Titan X Maxwell \textsc{gpu}).
A key advantage of our approach is that its compressed output can be decoded by any standard off-the-shelf \textsc{jpeg} implementation.  It serves to maintain the existing decoding complexity, the primary issue for distribution of electronic media.

Section 2 reviews \textsc{cnn} techniques used for object localization, semantic segmentation and class activation maps. We also discuss merits of using our technique over these methods.
Section 3 presents our new model which can generate a map showing multiple regions of interest. In Section 4 show how we combine this map to make \textsc{jpeg} semantically aware.
Sections 5 presents experimental results on a variety of image datasets and metrics. Section 6 concludes with future areas for research.


\longpage
\section{Review of localization using CNNs}

\textsc{cnn}s are multi-layered feed-forward architectures where the learned features at each level are the weights of the convolution filters to be applied to the output of the previous level. Learning is done via gradient-based optimization \cite{lecun1995convolutional}.
\textsc{cnn}s differ from fully connected neural networks in that the dimensions of the learned convolution filters are, in general, much smaller than the dimensions of the input image, so the learned features are forced to be localized in space. Also, the convolution operation uses the same weight kernel at every image location, so feature detection is spatially invariant.

Given an image $x$, and a convolution filter of size $n \times n$, then a convolutional layer performs the operation shown in equation~\ref{eqn_cnn}, where $\mathbf{W}$ is the learned filter.

\begin{equation}
    y_{ij} = \sum_{a=0}^{n} \sum_{b=0}^{n} \; \mathbf{W}_{ab} \; x_{(i+a)(j+b)}
    \label{eqn_cnn}
\end{equation}

In practice, multiple filters are learned in parallel within each layer, and thus the output of a convolution layer is a 3-\textit{d} feature map, where the depth represents the number of filters. The number of features in a given layer is a design choice, and may differ from layer to layer.
\textsc{cnn}s include a max pooling \cite{lecun1995convolutional} step after every or every other layer of convolution, in which the height and width of the feature map (filter response) are reduced by replacing several neighboring activations (coefficients), generally within a square window, with a single activation equal to the maximum within that window.  This pooling operation is strided, but the size of the pooling window can be greater than the stride, so windows can overlap.
This results in down-sampling of input data, and filters applied to such a map will have a larger receptive field (spatial support in the pixel space) for a given kernel size, thus reducing the number of parameters of the \textsc{cnn} model and allowing the training of much deeper networks.
This does not change the depth of the feature map, but only its width and height.
In practice, pooling windows are typically of size $2 \times 2$ or $4 \times 4$, with a stride of two, which reduces the number activations by 75\%.
\textsc{cnn}s apply some form of non-linear operation such as sigmoid $ (1-e^{-x})^{-1} $ or linear rectifier $max(0,x)$ on the output of each convolution operation.

Region-based \textsc{cnn}s use a moving window to maximize the posterior of the presence of an object \cite{girshick2015fast}.
Faster \textsc{rcnn}s \cite{ren2015faster} have been proposed, but they are still computationally expensive and are limited to determining the presence or absence of a single class of object within the entire image.
Moving-window methods are able to produce rectangular bounding boxes, but cannot produce object silhouettes.
In contrast, recent deep learning models proposed for semantic segmentation \cite{long2015fully}, \cite{girshick2014rich}, \cite{zheng2015conditional} are very good at drawing a close border around the objects.
However, these methods do not scale well to more than a small number of object categories (e.g. $20$) \cite{everingham2010pascal}.
Segmentation methods typically seek to produce a hard boundary for the object, in which every pixel is labeled as either part of the object or part of the background.
In contrast, class activation mapping produces a fuzzy boundary and is therefore able to capture pixels which are not part of any particular object, but are still salient to some spatial interaction between objects.
Segmentation techniques are also currently limited by the requirement for strongly-labeled data for training.
Obtaining training data where the locations of all the objects in the images are tagged is expensive and not scalable \cite{everingham2010pascal}. Our approach only requires image-level labels of object classes, without pixel-level annotation or bounding-box localization.

In a traditional \textsc{cnn}, there are two fully-connected (non-convolutional) layers as the final layers of the network. 
The final layer has one neuron for every class in the training data, and the final step in the inference is to normalize the activations of the last layer to sum to one.
The second to last layer, however, is fully connected to the last convolution layer, and a non-linearity is applied to its activations.
The authors of \cite{oquab2015object}, \cite{zhou2015learning} modify this second to last layer to allow for class localization.
In their architecture, the second to last layer is not learned, but consists of one neuron for each feature map, which has fixed  equally-weighted connections to each activation of its corresponding map.
No non-linearity is applied to the outputs of these neurons, so that each activation in this layer represents the global spatial average of one feature map from the previous layer.
Since the output of this layer is connected directly to the classification layer, each class will in essence learn a weight for each feature map from the final convolution layer.
Thus, given an image and a class, the classification weights for that class can be used to re-weight the layers of activations of the final convolution layer on that image.
These activations can be collapsed along the feature axis to create a class activation map, spatially localizing the best evidence for that class within that image. 
Figure~\ref{fig_comparison_of_map} (c) shows an example of such a map, the equation of which is given by

\longpage
\begin{equation}
    M_c(x,y) = \sum_{d \; \epsilon \; \mathbf{D}} w^{c}_{d} \; f_d(x,y)
    \label{eqn_gap}
\end{equation}

{\setlength{\parindent}{0cm} where $ w^{c}_{d} $ is the learned of class $c$ for feature map $d$.
Training for \textsc{cam} minimizes the cross entropy between objects' true probability distribution over classes (all mass given to the true class) and the predicted distribution, which is obtained as }

\begin{equation}
    P(c) = \frac{\text{exp}(\sum_{xy}{M_c(x,y)})}{\sum_c \text{exp}(\sum_{xy}{M_c(x,y)})}
    \label{eqn_posterior}
\end{equation}

Since \textsc{cam}s are trained to maximize posterior probability for the class, they tend to only highlight a single most prominent object.
This makes them useful for studying the output of \textsc{cnn}s, but not well suited to more general semantic saliency, as real world images  typically contain multiple objects of interest.

\longpage
\section{Multi-Structure Region of Interest}
We have developed a variant of \textsc{cam} which balances the activation for multiple objects and thus does not suffer from the issues of global average pooling.
Our method, \textit{Multi-Structure Region of Interest} (MS-ROI), allows us to effectively train on localization tasks independent of the number of classes.
For the purposes of semantic compression, obtaining a tight bound on the objects is not important. However, identifying and approximately locating all the objects is critical.
We propose a set of 3D feature maps in which each feature map is learned for an individual class, and is learned independently of the maps for other classes.
For $ \mathbf{L} $ layers, where each layer $l$ contains $d_l$ features, an image of size $ n \times n $, and with $\mathbf{C}$ classes, this results in a total activation size of
\longpage
\begin{equation*}
    \sum_{l \; \epsilon \; \mathbf{L}}{d_l \times \mathbf{C} \times \frac{n}{k^l} \times \frac{n}{k^l}}
\end{equation*}

where $k$ is the max pooling stride size.
This is computationally very expensive, and not practical for real world data.
\textsc{cnn}s designed for full-scale color images have many filters per layer and are several layers deep.
For such networks, learning a model with that many parameters would be unfeasible in terms of computational requirements.
We propose two techniques to make this idea feasible for large networks: 
(i) reduce the number of classes and increase the inter-class variance by combining similar classes, and  
(ii) share feature maps across classes to jointly learn lower level features.


Most \textsc{cnn} models are built for classification on the Large Scale Visual Recognition Challenge, commonly known as ImageNet.
ImageNet has one thousand classes and many of the classes are fine-grained delineations of various types of animals, flowers and other objects.
We significantly reduce the number of classes by collapsing these sets of similar classes to a single, more general class.
This is desirable because, for the purpose of selecting a class invariant `region of interest,' we do not care about the differences between Siberian husky and Eskimo dog or between Lace-flower and Tuberose.
As long as objects of these combined classes have similar structure and are within the same general category, the map produced will be almost identical.
Details of the combined classes used in our model are provided in the Experimental Results section.

It is self-evident that most images contain only a few classes and thus it is computationally inefficient to build a separate feature map for every class.
More importantly, many classes have similar lower-level features, even when the number of classes is relatively small.
The first few layers of a \textsc{cnn} learn filters which recognize small edges and motifs \cite{zeiler2014visualizing}, which are found across a wide variety of object classes.
Therefore, we propose parameter sharing across the feature maps for different classes.
This reduces the number of parameters and also allows for the joint learning of these shared, low-level features.

\begin{figure}
    \centering
    \includegraphics[scale=0.45]{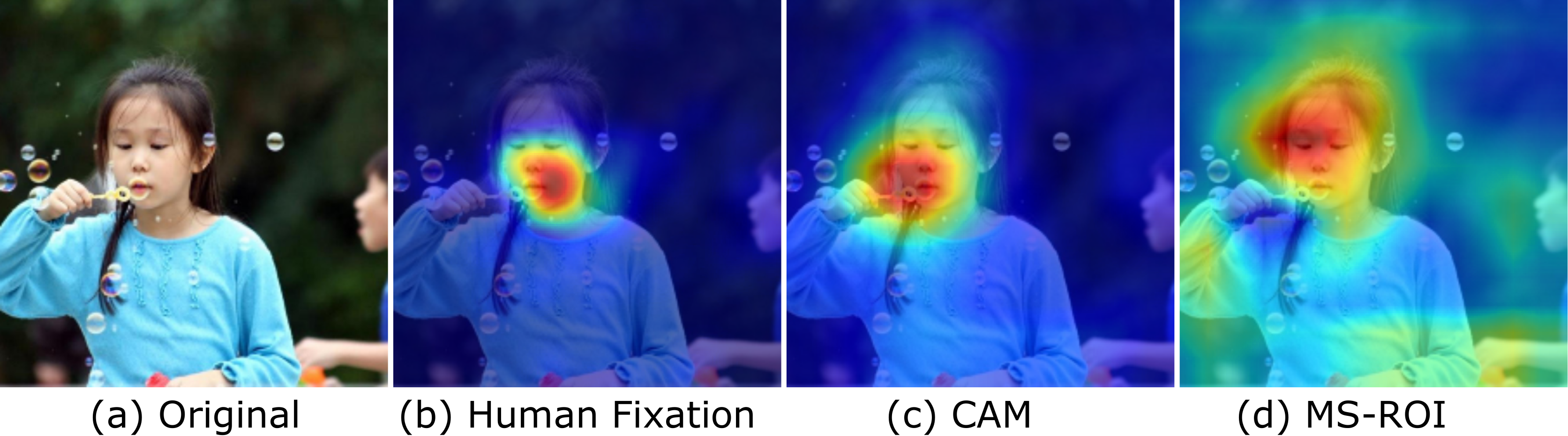}
    \caption{Comparison of various methods of detecting objects in an image \label{fig_comparison_of_map}}
\end{figure}

Although we do not restrict ourselves to a single most-probable class, it is desirable to eliminate the effects of activations for classes which are not present in the image.  
In order to accomplish this, we propose a thresholding operation which discards those classes whose learned features do not have a sufficiently large total activation when summed across all features and across the entire image.  
Let $Z^c_l$ denote the total sum of the activations of layer $\ell$ for all feature maps for a given class $c$.
Since our feature map is a $4$-dimensional tensor, $Z^c_l$ can be obtained by summation of this tensor over the three non-class dimensions.

\longpage
\begin{equation}
    Z^c_l = \sum_{d \; \epsilon \; \mathbf{D}} \sum_{x,y} f_d^c(x,y)
    \label{eqn_Zscore}
\end{equation}

Next, we use $Z^c_l$ to filter the classes. Computation of the multi-structure region of interest is shown below.

\begin{equation}
    \widehat{M}(x,y) = \sum_{c \; \epsilon \; \mathbf{c} }
    \begin{cases}
        \sum_d f_d^c(x,y), & \text{if}\ Z^c_l > T \\
        \phantom{this_is_empty} \\
      0 & \text{otherwise}
    \end{cases}
    \label{eqn_msroi}
\end{equation}

We use the symbol $ \widehat{M} $ to denote the multi-structure map generated by our proposed model in order to contrast it with the map generated using standard \textsc{cam} models, $ M $.
$ \widehat{M} $ is a sum over all classes with total activations $Z^c_l$ beyond a threshold value $T$.
$T$ is determined during the training or chosen as a hyper-parameter for learning. 
In practice, it is sufficient to \textit{argsort} $Z^c_l$ and pick the top five classes and combine them via a sum weighted by their rank. 
It should be noted that, because $ \widehat{M} $ is no longer a reflection of the class of the image, we use the term `region of interest'.

A comparison of our model (MS-ROI) with \textsc{cam} and human fixation is shown in Figure~\ref{fig_comparison_of_map}. 
Only our model identifies the face of the boy on the right as well the hands of both children at the bottom. 
When doing compression, it is important that we do not lower the quality of body extremities or other objects which other models may not identify as critical to the primary object class of the image.
If a human annotator were to paint the areas which should be compressed at better quality, we believe the annotated area would be closer to that captured by our model.

To train the model to maximize the detection of all objects, instead of using a softmax function as in equation~\ref{eqn_posterior}, we use sigmoid, which does not marginalize the posterior over the classes.
Thus the likelihood of a class $c$ is given by equation~\ref{eqn_our_posterior}. 

\begin{equation}
    P(c) = \frac{1}{1 + \text{exp}(Z^c_l)}
    \label{eqn_our_posterior}
\end{equation}

\longpage
\section{Integrating MS-ROI map with JPEG}
We obtain from MS-ROI a saliency value for each pixel in the range [0,1], where 1 indicates maximum saliency.
Then discretize these saliency values into $k$ levels, where $k$ is a tune-able hyper-parameter.
The lowest level contains pixels of saliency $[0,1/k]$, the second level contains pixels of saliency $(1/k,2/k]$ and so forth.
We next select a range of \textsc{jpeg} quality levels, $Q_l$ to $Q_h$.  Each saliency level will be compressed using a $Q$ value drawn from this range, corresponding to that level.
In other words, saliency level $n$, with saliency range $[n/k,(n+1)/k]$ will be compressed using 

\begin{equation}
Q_n = Q_l + \frac{n*(Q_h - Q_l)}{k}
\end{equation}


For each level $l \le n \le h$, we obtain a decoded \textsc{jpeg} of the image after encoding at quality level $Q_n$.
For each $8 \times 8$ block of our output image, we select the block of color values obtained by the \textsc{jpeg} corresponding to that block's saliency level.  This mosiac of blocks is finally compressed using a standard \textsc{jpeg} encoder with the desired output quality to produce a file which can be decoded by any off-the-shelf \textsc{jpeg} decoder.

Details of our choices for $k$, $Q_l$ and $Q_h$, as well as target image sizes are provided in the next section.  A wider range of $Q_l$ and $Q_h$ will tend to produce stronger results, but at the expense of very poor quality in non-salient regions.

\longpage
\section{Experimental Results}
We trained our model with the Caltech-256 dataset \cite{griffin2007caltech}, which contains 256 classes of man-made and natural objects, common plants and animals, buildings, etc.
We believe this offers a good balance between covering more classes as compared to CIFAR-100 which contains only 100 classes, and avoiding overly finely-grained classes as in ImageNet with 1000 classes \cite{imagenet_cvpr09}.
For the results reported here, we experimented with several stacked layers of convolution as shown in the diagram below:
\begin{equation*}
    \text{IMAGE} \longmapsto \bigg[ \big[ \text{CONV} \rightarrow \text{RELU}\big]^2 \rightarrow \text{MAXPOOL} \bigg]^5 \longmapsto \text{MS-ROI} \longmapsto \text{MAP}
\end{equation*}

$\text{MS-ROI}$ refers to the operation shown in the equation~\ref{eqn_msroi}.
To obtain the final image we discretize the heat-map into five levels and use \textsc{jpeg} quality levels $Q$ in increments of ten from $Q_l=30$ to $Q_h=70$.
For all experiments, the file size of the standard \textsc{jpeg} image and the \textsc{jpeg} obtained from our model were kept within $\pm1\%$ of each other.
On average, salient regions were compressed at $Q_f=65$, and non-salient regions were compressed at $Q=45$.
The overall $Q$ for the final image generated using our model was $Q=55$, whereas for all standard \textsc{jpeg} samples, $Q$ was chosen to be 50.

\begin{figure}
    \centering
    \includegraphics[scale=0.37]{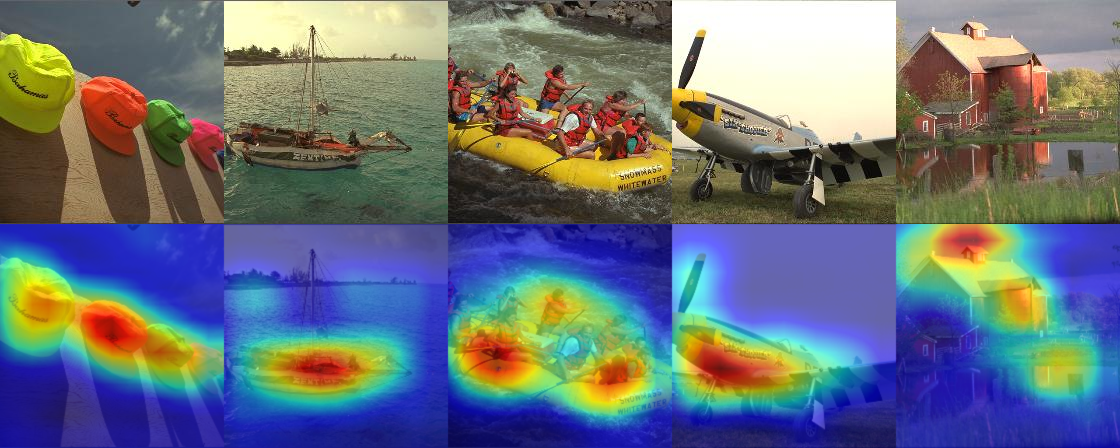}
    \caption{Sample of our map for five KODAK images.\label{fig_kodak_combo}}
\end{figure}

\begin{figure}
    \centering
    \includegraphics[scale=0.23]{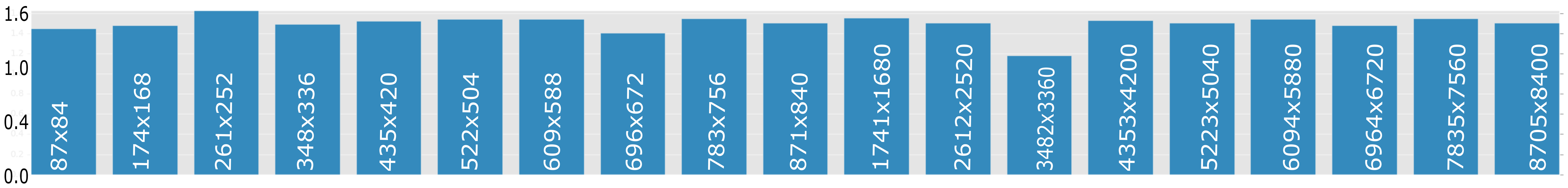}
    \caption{\texttt{PSNR-HVS} of our model - \textsc{jpeg} across various image size (higher is better). \label{fig_size}}
\end{figure}

\begin{figure}
    \centering
    \includegraphics[scale=0.23]{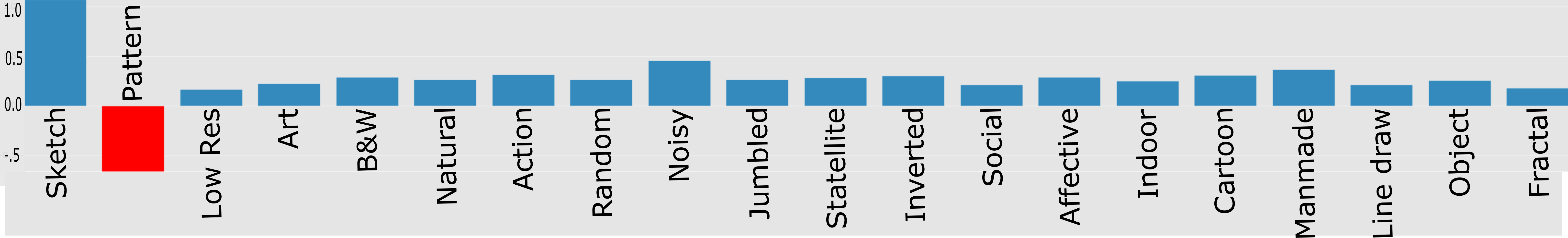}
    \caption{\texttt{PSNR-HVS} of our model - \textsc{jpeg} across various categories of MIT Saliency dataset (higher is better). \label{fig_mit}}
\end{figure}

We tested on the Kodak PhotoCD set (24 images) and the the MIT dataset (2,000 images).
Kodak is a well known dataset consisting primarily of natural outdoor color images.
Figure~\ref{fig_kodak_combo} shows a sample of five of these images, along with the corresponding heatmaps generated by our algorithm; the first four show typical results which strongly capture the salient content of the images, while the fifth is a rare case of partial failure, in which the heatmap does not fully capture all salient regions.
The MIT set allows us to compare results across twenty categories. In Table~\ref{tbl_results} we only report averaged results across `Outdoor Man-made' and `Outdoor Natural' categories (200 images), as these categories are likely to contain multiple semantic objects, and are therefore appropriate for our method.
Both datasets contain images of smaller resolutions, but the effectiveness of perceptual compression is more pronounced for larger images.
Therefore, we additionally selected a very large image of resolution $8705\times8400$, which we scale to a range of sizes to demonstrate the effectiveness of our system at a variety of resolutions.
See Figure~\ref{fig_size} for the image sizes used in this experiment.
Both Figure~\ref{fig_size} and Figure~\ref{fig_mit} show the \texttt{PSNR-HVS} difference between our model and standard \textsc{jpeg}. Positive values indicate our model has higher performance compared to standard \textsc{jpeg}.
In addition to an array of standard quality metrics, we also report a \texttt{PSNR} value calculated only for those regions our method has identified as salient, which we term \texttt{PSNR-S}.
By examining only regions of high semantic saliency, this metric demonstrates that our compression method is indeed able to preserve visual quality in targeted regions, without sacrificing performance on traditional image-level quality metrics or compression ratio.
It should be noted that the validity of this metric is dependent on the correctness of the underlying saliency map, and thus should only be interpreted to demonstrate the success of the final image construction in preserving details highlighted by that map.

\begin{table}
\footnotesize
\centering
\begin{tabular}{lccccccc}
			 \toprule
               & \texttt{PSNR-S} \phantom{so} & \texttt{PSNR} \phantom{so} & \texttt{PSNR-HVS} \phantom{so}& \texttt{PSNR-HVSM} \phantom{so}& \texttt{SSIM} \phantom{so}& \texttt{MS-SSIM} \phantom{so}& \texttt{VIFP}\phantom{so} \\
              \midrule
              & \multicolumn{6}{c}{Kodak PhotoCD [24 images]}                           \\
              \midrule
Std \textsc{jpeg} &   33.91   & 34.70     & 34.92          & 42.19  & 0.969  & 0.991 & 0.626      \\
			  \cline{2-8}
Our model &  39.16    & 34.82    & 35.05          & 42.33  & 0.969  & 0.991 & 0.629     \\
			  \midrule
              & \multicolumn{6}{c}{MIT Saliency Benchmark [Outdoor Man-made + Natural, 200 images]}              \\
			  \midrule
Std \textsc{jpeg} & 36.9 & 31.84    & 35.91   & 45.37 & 0.893 & 0.982     & 0.521      \\
			  \cline{2-8}
Our model & 40.8 & 32.16   & 36.32   & 45.62 &  0.917 & 0.990      & 0.529      \\
			  \midrule
              & \multicolumn{6}{c}{Re-sized images of a very large image, see fig:~\ref{fig_size} [20 images]}          \\
			  \midrule
Std \textsc{jpeg} & 35.4   & 27.46  & 33.12 & 43.26     & 0.912 & 0.988 & 0.494     \\
			  \cline{2-8}
Our model & 39.6  & 28.67  & 34.63 & 44.89     & 0.915 & 0.991 & 0.522     \\
			  \toprule
\end{tabular}
\caption{Results across datasets \label{tbl_results}}
\end{table}
        
\longpage

The results in Table~\ref{tbl_results} show the success of our method in maintaining or improving performance on traditional image quality metrics.
Further, given the efficacy of our method in identifying multiple regions of interest, the \texttt{PSNR-S} measurements demonstrate the power of our method to produce superior visual quality in subjectively important regions.


    Figure~\ref{fig_mit} shows the performance of our model across all categories of the MIT dataset.
    Performance was strongest in categories like `Outdoor Natural', `Outdoor Man Made', `Action' and `Object', while categories like `Line Drawing', `Fractal' and `Low Resolution' showed the least improvement.
    Not surprisingly, the category `Pattern', which lacks semantic objects, is the only category where our model did not improve upon standard \textsc{jpeg}.
    Figure~\ref{fig_size} shows results on the same image scaled to different sizes. 
    Because our model benefits from the scale-invariance of \textsc{cnn}s, we are able to preserve performance across a wide range of input sizes.

\longpage

\section{Conclusion and Future research}
We have presented a model which can learn to detect multiple objects at any scale and generate a map of multiple semantically salient image regions.
This provides sufficient information to perform variable-quality image compression, without providing a precise semantic segmentation.
Unlike region-based models, our model does not have to iterate over many windows.
We sacrifice exact localization for the ability to detect multiple salient objects.
Our variable compression improves upon visual quality without sacrificing compression ratio. 
Encoding requires a single inference over the pre-trained model, the cost of which is reasonable when performed using a \textsc{gpu}, along with a standard \textsc{jpeg} encoder. The cost of decoding, which employs a standard, off-the-shelf \textsc{jpeg} decoder remains unchanged.
We believe it will be possible to incorporate our approach into other lossy compression methods such as \textsc{jpeg} 2000 and vector quantization, a subject of future work.
Improvements to the power of our underlying \textsc{cnn}, addressing evolving visual quality metrics, and other applications such as video compression, are also potential areas of future work.


\bibliographystyle{ieeetr}
{\footnotesize \bibliography{refs}}

\begin{thebibliography}{10}

\bibitem{ginesu2012objective}
G.~Ginesu, M.~Pintus, and D.~D. Giusto, ``Objective assessment of the webp
  image coding algorithm,'' {\em Signal Processing: Image Communication},
  vol.~27, no.~8, pp.~867--874, 2012.

\bibitem{toderici2016full}
G.~Toderici, D.~Vincent, N.~Johnston, S.~J. Hwang, D.~Minnen, J.~Shor, and
  M.~Covell, ``Full resolution image compression with recurrent neural
  networks,'' {\em arXiv preprint arXiv:1608.05148}, 2016.

\bibitem{jiang2015salicon}
M.~Jiang, S.~Huang, J.~Duan, and Q.~Zhao, ``Salicon: Saliency in context,'' in
  {\em 2015 IEEE Conference on Computer Vision and Pattern Recognition (CVPR)},
  pp.~1072--1080, IEEE, 2015.

\bibitem{kerofsky2015perceptual}
L.~Kerofsky, R.~Vanam, and Y.~Reznik, ``Perceptual adaptation of objective
  video quality metrics,'' in {\em Proc. Ninth International Workshop on Video
  Processing and Quality Metrics (VPQM)}, 2015.

\bibitem{richter2009ms}
T.~Richter and K.~J. Kim, ``A ms-ssim optimal jpeg 2000 encoder,'' in {\em 2009
  Data Compression Conference}, pp.~401--410, IEEE, 2009.

\bibitem{richter2011ssim}
T.~Richter, ``Ssim as global quality metric: a differential geometry view,'' in
  {\em Quality of Multimedia Experience (QoMEX), 2011 Third International
  Workshop on}, pp.~189--194, IEEE, 2011.

\bibitem{ssim}
Z.~Wang, A.~C. Bovik, H.~R. Sheikh, and E.~P. Simoncelli, ``Image quality
  assessment: from error visibility to structural similarity,'' {\em IEEE
  transactions on image processing}, vol.~13, no.~4, pp.~600--612, 2004.

\bibitem{msssim}
Z.~Wang, E.~P. Simoncelli, and A.~C. Bovik, ``Multiscale structural similarity
  for image quality assessment,'' in {\em Signals, Systems and Computers, 2004.
  Conference Record of the Thirty-Seventh Asilomar Conference on}, vol.~2,
  pp.~1398--1402, Ieee, 2003.

\bibitem{vifp}
H.~R. Sheikh and A.~C. Bovik, ``Image information and visual quality,'' {\em
  IEEE Transactions on Image Processing}, vol.~15, no.~2, pp.~430--444, 2006.

\bibitem{psnrhvs}
K.~Egiazarian, J.~Astola, N.~Ponomarenko, V.~Lukin, F.~Battisti, and M.~Carli,
  ``New full-reference quality metrics based on hvs,'' in {\em CD-ROM
  proceedings of the second international workshop on video processing and
  quality metrics, Scottsdale, USA}, vol.~4, 2006.

\bibitem{psnrhvsm}
N.~Ponomarenko, F.~Silvestri, K.~Egiazarian, M.~Carli, J.~Astola, and V.~Lukin,
  ``On between-coefficient contrast masking of dct basis functions,'' in {\em
  Proceedings of the third international workshop on video processing and
  quality metrics}, vol.~4, 2007.

\bibitem{krizhevsky2012imagenet}
A.~Krizhevsky, I.~Sutskever, and G.~E. Hinton, ``Imagenet classification with
  deep convolutional neural networks,'' in {\em Advances in neural information
  processing systems}, pp.~1097--1105, 2012.

\bibitem{zeiler2014visualizing}
M.~D. Zeiler and R.~Fergus, ``Visualizing and understanding convolutional
  networks,'' in {\em European Conference on Computer Vision}, pp.~818--833,
  Springer, 2014.

\bibitem{he2015deep}
K.~He, X.~Zhang, S.~Ren, and J.~Sun, ``Deep residual learning for image
  recognition,'' in {\em The IEEE Conference on Computer Vision and Pattern
  Recognition (CVPR)}, June 2016.

\bibitem{girshick2014rich}
R.~Girshick, J.~Donahue, T.~Darrell, and J.~Malik, ``Rich feature hierarchies
  for accurate object detection and semantic segmentation,'' in {\em
  Proceedings of the IEEE conference on computer vision and pattern
  recognition}, pp.~580--587, 2014.

\bibitem{kummerer2014deep}
M.~K{\"u}mmerer, L.~Theis, and M.~Bethge, ``Deep gaze i: Boosting saliency
  prediction with feature maps trained on imagenet,'' {\em arXiv preprint
  arXiv:1411.1045}, 2014.

\bibitem{mnih2014recurrent}
V.~Mnih, N.~Heess, A.~Graves, {\em et~al.}, ``Recurrent models of visual
  attention,'' in {\em Advances in Neural Information Processing Systems},
  pp.~2204--2212, 2014.

\bibitem{zund2013content}
F.~Z{\"u}nd, Y.~Pritch, A.~Sorkine-Hornung, S.~Mangold, and T.~Gross,
  ``Content-aware compression using saliency-driven image retargeting,'' in
  {\em 2013 IEEE International Conference on Image Processing}, pp.~1845--1849,
  IEEE, 2013.

\bibitem{stella2009image}
X.~Y. Stella and D.~A. Lisin, ``Image compression based on visual saliency at
  individual scales,'' in {\em International Symposium on Visual Computing},
  pp.~157--166, Springer, 2009.

\bibitem{dai2016r}
J.~Dai, Y.~Li, K.~He, and J.~Sun, ``R-fcn: Object detection via region-based
  fully convolutional networks,'' {\em arXiv preprint arXiv:1605.06409}, 2016.

\bibitem{liu2015predicting}
N.~Liu, J.~Han, D.~Zhang, S.~Wen, and T.~Liu, ``Predicting eye fixations using
  convolutional neural networks,'' in {\em Proceedings of the IEEE Conference
  on Computer Vision and Pattern Recognition}, pp.~362--370, 2015.

\bibitem{lecun1995convolutional}
Y.~LeCun and Y.~Bengio, ``Convolutional networks for images, speech, and time
  series,'' {\em The handbook of brain theory and neural networks}, vol.~3361,
  no.~10, p.~1995, 1995.

\bibitem{girshick2015fast}
R.~Girshick, ``Fast r-cnn,'' in {\em Proceedings of the IEEE International
  Conference on Computer Vision}, pp.~1440--1448, 2015.

\bibitem{ren2015faster}
S.~Ren, K.~He, R.~Girshick, and J.~Sun, ``Faster r-cnn: Towards real-time
  object detection with region proposal networks,'' in {\em Advances in neural
  information processing systems}, pp.~91--99, 2015.

\bibitem{long2015fully}
J.~Long, E.~Shelhamer, and T.~Darrell, ``Fully convolutional networks for
  semantic segmentation,'' in {\em Proceedings of the IEEE Conference on
  Computer Vision and Pattern Recognition}, pp.~3431--3440, 2015.

\bibitem{zheng2015conditional}
S.~Zheng, S.~Jayasumana, B.~Romera-Paredes, V.~Vineet, Z.~Su, D.~Du, C.~Huang,
  and P.~H. Torr, ``Conditional random fields as recurrent neural networks,''
  in {\em Proceedings of the IEEE International Conference on Computer Vision},
  pp.~1529--1537, 2015.

\bibitem{everingham2010pascal}
M.~Everingham, L.~Van~Gool, C.~K. Williams, J.~Winn, and A.~Zisserman, ``The
  pascal visual object classes (voc) challenge,'' {\em International journal of
  computer vision}, vol.~88, no.~2, pp.~303--338, 2010.

\bibitem{oquab2015object}
M.~Oquab, L.~Bottou, I.~Laptev, and J.~Sivic, ``Is object localization for
  free?-weakly-supervised learning with convolutional neural networks,'' in
  {\em Proceedings of the IEEE Conference on Computer Vision and Pattern
  Recognition}, pp.~685--694, 2015.

\bibitem{zhou2015learning}
B.~Zhou, A.~Khosla, A.~Lapedriza, A.~Oliva, and A.~Torralba, ``Learning deep
  features for discriminative localization,'' in {\em The IEEE Conference on
  Computer Vision and Pattern Recognition (CVPR)}, June 2016.

\bibitem{griffin2007caltech}
G.~Griffin, A.~Holub, and P.~Perona, ``Caltech-256 object category dataset,''
  2007.

\bibitem{imagenet_cvpr09}
J.~Deng, W.~Dong, R.~Socher, L.-J. Li, K.~Li, and L.~Fei-Fei, ``Imagenet: A
  large-scale hierarchical image database,'' in {\em Computer Vision and
  Pattern Recognition, 2009. CVPR 2009. IEEE Conference on}, pp.~248--255,
  IEEE, 2009.

\end{thebibliography}
\end{document}